\documentclass[amsmath,amsfonts,amssymb,twocolumn,superscriptaddress,showpacs]{revtex4}

\usepackage{graphicx,psfrag}
\usepackage{dcolumn}
\usepackage{bm}
\usepackage{mathbbol}

\newcommand{\avg}[1]{\langle #1 \rangle}

\begin{document}

\title{\mbox{Generalized Coherent States as Preferred States
of Open Quantum Systems}}

\author{Sergio Boixo} 
\affiliation{\mbox{Department of Physics and
Astronomy, University of New Mexico, Albuquerque, NM 87131, USA}}
\affiliation{\mbox{CCS-3: Modeling, Algorithms, and Informatics, B256,
Los Alamos National Laboratory, Los Alamos, NM 87545, USA}}

\author{Lorenza Viola}
\affiliation{\mbox{Department of Physics and Astronomy, Dartmouth
College, 6127 Wilder Laboratory, Hanover, NH 03755, USA}}

\author{Gerardo Ortiz}
\affiliation{Department of Physics, Indiana University, Bloomington,
IN 47405, USA}


\begin{abstract}
  We investigate the connection between {\em quasi-classical (pointer)
  states} and {\em generalized coherent states} (GCSs) within an
  algebraic approach to Markovian quantum systems (including
  bosons, spins, and fermions).  We establish conditions for the GCS
  set to become most {\it robust} by relating the rate of purity loss
  to an invariant measure of uncertainty derived from quantum
  Fisher information.  We find that, for damped bosonic modes, the
  stability of canonical coherent states is confirmed in a variety of
  scenarios, while for systems described by (compact) Lie algebras
  stringent symmetry constraints must be obeyed for the GCS set to be
  preferred.  The relationship between GCSs, minimum-uncertainty
  states, and decoherence-free subspaces is also elucidated.
\end{abstract}

\maketitle


The quest for quantum states resembling classical behavior dates back
to the early days of Quantum Mechanics, when Schr\"odinger identified
the closest to classical states of a quantum harmonic oscillator
\cite{Schroedinger26}.  Following the work by Glauber and Sudarshan
\cite{Roy63,Sud63}, canonical coherent states (CCSs) played a
pervasive role across photon and atom optics ever since
\cite{Scully}. The distinctive properties enjoyed by CCSs are rooted
in the algebraic structure of the harmonic oscillator Hamiltonian.
For systems described by different algebras, CCSs are replaced by {\em
generalized coherent states} (GCSs) \cite{Gil}. Like CCSs, GCSs are
{\em minimum uncertainty states} \cite{Del}, admit a natural
phase-space structure \cite{Gil}, and are temporally stable under
Hamiltonian evolution \cite{Tstab}. GCSs are ubiquitous in Nature:
Aside from optical and atomic physics, states of matter such as BCS
superconductors or normal Fermi liquids, for instance, are typically
described by GCSs \cite{Gil}.  Recently, GCSs have been characterized
as un-entangled within the framework of generalized entanglement
\cite{GE} -- such a property being related to efficient simulatability
of Lie-algebraic models of quantum computation \cite{Somma06}.

While all of the above properties make GCSs excellent candidates for
{\em quasi-classical} states, no real-world system is isolated from
its environment, causing pure quantum states to rapidly deteriorate
into mixtures.  How can quantum properties survive in the macroworld?
Acknowledging the role of the environment and characterizing
classicality as an emergent property of {\em open} quantum systems
lies at the heart of the decoherence program \cite{Zurek03}. {\em
Pointer states} (PSs), in particular, are distinguished by their
ability to persist in spite of the environment, thus are the {\it
preferred} states in which open systems are found in practice.  PSs
exhibit {\em minimum purity loss} \cite{Zurek1993}. Several authors
confirmed that for harmonic systems CCSs emerge under fairly generic
conditions \cite{Old}, which naturally raises some fundamental
questions: What is the relationship between minimum purity loss and
minimum uncertainty?  Do the GCS and PS sets coincide in general?  To
what extent can stability properties against decoherence explain the
special status that GCSs have in Physics?

In this Letter, we address the above questions for Markovian quantum
systems \cite{BP02} described by a Lie algebra ${\mathfrak g}$.  Using
the concept of {\em quantum Fisher information} (QFI) from estimation
theory, we derive an invariant measure of uncertainty, and establish
under which conditions such a measure is proportional to the dynamical
rate of purity loss -- thereby obtaining a streamlined method to link
PSs to GCSs of ${\mathfrak g}$.  Besides extending earlier results on
single- and multi-mode bosonic algebras, we devote special emphasis to
({semisimple compact}) Lie algebras describing quantum spins and
fermion systems \cite{Gil}. For irreducible representations (irreps),
we find that more restrictive symmetry requirements than in the
bosonic case must be obeyed for the full GCS set to be preferred.  For
reducible representations, we propose an explicit characterization of
the generalized minimum-uncertainty states manifold, which allows us
to compare with information preserving structures as introduced in
quantum information science. While decoherence-free subspaces
(DFSs)~\cite{dfs} are found to be natural multi-dimensional extensions
of PSs and GCSs under appropriate conditions, genuine noiseless
subsystems (NSs)~\cite{KLV} are not -- further pointing to the
distinction between preserved pure states and preserved information
which is quintessential to the NS idea.

{\em Dynamical-algebraic setting and purity loss.}-- Let $S$
denote the open system of interest, defined on a (separable) Hilbert
space ${\cal H}$. We assume that the state $\rho(t)$
of $S$ evolves according to a quantum Markovian master equation, that
is, in units $\hbar =1$ \cite{BP02},
\begin{eqnarray}
&& \dot{\rho}(t) = {\cal L}[\rho (t)] = -i [H_S, \rho(t)] + {\cal
D}[\rho (t)] \,, \nonumber \\ && {\cal D}[\rho] = \frac{1}{2}
\sum_\ell \Big( [L_\ell \rho, L_\ell^\dagger] + [L_\ell, \rho
L_\ell^\dagger] \Big)\,,
\label{lme}
\end{eqnarray}
where ${\cal L}[\rho]$, ${\cal D}[\rho]$ denote the Lindblad generator
and dissipator, respectively. Both the physical (renormalized)
Hamiltonian $H_S$ and the Lindblad operators $L_\ell$ are bounded and
time-independent. In the so-called weak coupling limit (WCL), the
perturbation due to ${\cal D}[\rho]$ is assumed to be weak enough for
the relaxation time $\tau_R$ to set the slowest characteristic time
scale -- both relative to the reservoir correlation time, $\tau_R \gg
\tau_c$ (Born-Markov condition), and the system dynamical time scale,
$\tau_R \gg \tau_S$ (enabling the rotating wave approximation to
apply). The WCL effectively constrains the Lindblad operators to
obey the condition 
\begin{equation}
 [H_S, L_\ell] =  \lambda_\ell L_\ell\,,
\;\; \forall \ell\,,
\label{wcl}
\end{equation}
where $\lambda_\ell \in {\mathbb R}$ are related to the Bohr
frequencies of $H_S$ \cite{BP02}. Because the WCL is the only rigorous
constructive approach for arbitrary temperature, primary emphasis will
be given here to evolutions obeying Eq.~(\ref{wcl}).  In such a case,
we define the {\em dynamical Lie algebra ${\mathfrak g}$ associated to
$S$} as the minimal Lie algebra generated by the $L_\ell$.  In
phenomenological applications (and also the high-temperature limit
and/or non-stationary environments), (\ref{wcl}) is often waived.
When considering such cases, we explicitly include $H_S$ among the
generators of ${\mathfrak g}$.
 
PSs are quantum states suffering minimal decoherence over a range of
dynamical time scales \cite{Zurek1993}.  They may be found by
minimizing the {\em average purity loss},
\begin{eqnarray}
\overline{\Pi}_{|\psi\rangle}\equiv \frac{\Delta \Pi (\tau)}{\tau}=
\frac{1}{\tau} \int_0^\tau dt \, \dot{\Pi}_{|\psi\rangle}(t) \,,
\label{sieve}
\end{eqnarray}
with $\Pi_{|\psi\rangle}(t)= 1-\mbox{Tr}\{\rho_{|\psi\rangle}^2
(t)\}$, $\rho_{|\psi\rangle}(t)$ satisfying Eq.~(\ref{lme}),
$\rho_{|\psi\rangle}(0)$ $= |\psi\rangle\langle\psi|$, and $\tau >0$
to be determined. By moving to the Heisenberg picture, the rate of
purity loss becomes
\begin{equation}
\dot{\Pi}_{|\psi\rangle}(t)= 2\sum_\ell (\Delta L_\ell (t))^2 \, ,
\label{loss}
\end{equation}
where, for a generic operator $O$, the {\em (quasi)variance} is
$(\Delta O)^2_{|\psi\rangle} = || (O -\langle O\rangle_{|\psi\rangle}
) |\psi\rangle ||^2 \equiv (\Delta O)^2$.

In the WCL, it is legitimate to treat the effects of ${\cal D}$ to
first order in time.  By exploiting Eq.~(\ref{wcl}), this yields
\begin{equation}
L_\ell(t)=e^{t {\cal L}^\dagger} [L_\ell] \approx e^{it H_S} L_\ell
e^{-it H_S} =e^{it\lambda_\ell} L_\ell\, ,
\label{Lt}
\end{equation}
and the (first order) rate of purity loss becomes {\em
time-independent} and equal to the average purity loss:
\begin{equation}
\dot{\Pi}_{|\psi\rangle} = \overline{\Pi}_{|\psi\rangle} = 2\sum_\ell
(\Delta L_\ell)^2 \,.
\label{wcl2}
\end{equation}


{\em GCSs and invariant uncertainty.}-- CCSs correspond to the action
of the so-called Heisenberg-Weyl group on the single-mode Fock vacuum:
$|\eta\rangle = {\tt D}(\eta) |0\rangle$, where ${\tt D}(\eta)=
\exp(\eta a^\dagger -\eta^\ast a)$ denotes the phase-space
displacement operator constructed from the oscillator algebra
$\mathfrak{h}_3=\{ \openone, a, a^\dagger \}$. GCSs are a
generalization of this construction specified by three inputs
\cite{Gil}: a dynamical Lie group ${\cal G}$, with associated Lie
algebra ${\mathfrak g}$~\cite{caveat}; a unitary irrep $\Gamma$ of
${\cal G}$; and a normalized reference state
$|\Lambda\rangle$. Following \cite{Gil}, we require $|\Lambda\rangle$
to have {\em maximum symmetry}~\cite{symmetry}.  The GCSs associated
to $({\cal G}, \Gamma, |\Lambda\rangle)$ are defined as ``generalized
displacements'' of $|\Lambda\rangle$: $|\Lambda,\eta\rangle= {\tt
D}(\{\eta\}) |\Lambda\rangle$, $ {\tt D}(\{\eta\}) \in {\cal
G}$. Among GCSs of finite-dimensional semisimple Lie algebras,
noteworthy examples include $\mathfrak{su}(2)$-spin GCSs (also known
as atomic coherent states~\cite{Arecchi}) as well as $N$-fermion GCSs
associated to $\mathfrak{so}(2N) =\mbox{span}\{ c^\dagger_i c^{\;}_j,
c^\dagger_i c^\dagger_j , c^{\;}_i c^{\;}_j \,|\, 1 \leq i,j \leq N\}$
(including, as mentioned, the ground state of BCS theory~\cite{Gil}).

It is well known that CCSs achieve the lower bound in the Heisenberg
uncertainty, $\Delta x\Delta p \geq 1/2$ (in appropriate units).  {\em
Generalized uncertainties} set limits on the precision with which a
given parameter may be estimated~\cite{Braunstein1996}. Such a
precision is quantified using the QFI~\cite{hayashi}.  A parameter
$\theta$ has generator $K_\theta$ if $\partial_\theta \rho_\theta =
-i[K_\theta,\rho_\theta]$, $\rho_\theta$ being the quantum state.  If
$\rho_\theta$ is pure, the corresponding QFI is $\mathcal{I}_\theta =
4 (\Delta K_\theta)^2$, and the Cramer-Rao inequality for unbiased
estimation dictates that
\begin{equation}
\delta \theta^2 \cdot \mathcal{I}_\theta = \delta \theta^2 \cdot 4
(\Delta K_\theta)^2 \ge 1 \;.
\end{equation}
With $\theta = x$, $K_\theta = p$, the above standard Heisenberg
uncertainty principle is recovered.  With $\theta=t$, $K_\theta = H$,
we obtain the time-energy uncertainty, whereas $\theta=\phi$,
$K_\theta = n$ gives the phase-number uncertainty.

The extension of these ideas to multiparameter estimation becomes
relevant for the purpose of connecting PSs to GCSs.  Let $\vec \theta
= (\theta_1, \dots, \theta_n)$. For pure states, the QFI matrix ${\cal
I}_{\vec \theta}$ is 
\begin{equation}
   [\mathcal I_{\vec \theta}]_{j,k} = 4 \avg
  {(K_{\vec\theta_j} - \avg{K_{\vec\theta_j}})(K_{\vec\theta_k} -
  \avg{K_{\vec\theta_k}})}\; .
\end{equation}
Then the quantity ${\rm tr}\, {\cal I}_{\vec \theta}/4=(\Delta
\mathcal I)^2$ represents a {\em scalar} measure of uncertainty.  The
quantum Cramer-Rao inequality also holds in the multiparameter
setting. For example, with $K_{\theta_1}=x,K_{\theta_2}=p$, $(\Delta
\mathcal I)^2 =(\Delta x)^2 + (\Delta p)^2$, which is minimized by
CCSs (but not squeezed states). If the generators $K_{\vec \theta}$
form a normalized basis of $\mathfrak{g}$ (with respect to the Killing
metric), we recover the {\em invariant uncertainty} identified by
Delbourgo~\cite{Del}. By a suitable change of basis, $(\Delta {\cal
I})^2$ may always be expressed as a sum of variances of observables, $
(\Delta \mathcal I)^2 = \sum_j (\Delta X_j)^2$, with $X_j^\dagger =
X_j$.  Physically, $(\Delta {\cal I})^2$ quantifies the {\em global}
amount of uncertainty of $|\psi\rangle$ in a way which is invariant
under arbitrary transformations in ${\cal G}$.  The following {\em
state-independent} lower bound for $(\Delta {\cal
I})^2_{|\psi\rangle}$ holds:

\vspace*{0.5mm}

{\bf Theorem 1} ({\em Invariant uncertainty principle}): {Let $\Gamma$
be an irrep of a semisimple Lie algebra ${\mathfrak g}$ with
highest-weight element $|\Lambda\rangle$ (a reference state for the
GCSs of $\mathfrak g$). Then
\begin{equation}
  (\Delta {\cal I})^2_{|\psi\rangle} \geq  \sum_j k_j \langle
  \alpha_j,\alpha_j\rangle\,,
\label{invunc2}
\end{equation}
\noindent 
where $\Lambda =\sum_j k_j \alpha_j$ in terms of simple roots
$\alpha_j$. The lower bound is attained iff $|\psi\rangle$ is a GCS of
$\mathfrak{g}$.}

\vspace*{0.5mm}

A complete proof of Theorem 1 is rather technical, and while a sketch
is included in \cite{Proof1}, we refer to \cite{Long} for full detail.
Physically, notice that for a spin-$J$ irrep of $\mathfrak{su}(2)$,
Theorem 1 recovers the familiar uncertainty relation for angular
momentum GCSs, $(\Delta J)^2 =\sum_{a=x,y,z} (\Delta J_a)^2 \geq
J$. Although $\mathfrak{h}_3$ is {\em not} semisimple, it is
intriguing that invariant uncertainty relationships structurally
similar to (\ref{invunc2}) emerge upon mapping $\mathfrak{su}(2)$ into
$\mathfrak{h}_3$ via a standard Holstein-Primakoff transformation in
the large-$J$ limit, $(\Delta J)^2/J \mapsto (\Delta a)^2 +(\Delta
a^\dagger)^2 = (\Delta x)^2 +(\Delta p)^2\geq 1$.

\vspace*{0.5mm}

{\em GCSs vs PSs for irreducibly represented algebras.}-- We first
derive the conditions for the emergence of multi-mode CCSs in damped
bosonic systems \cite{Scully}, with $H_S =\sum_{i=1}^n \omega_i
(a^\dagger_i a_i^{\;} + 1/2)$, $\omega_i >0$.

\vspace*{0.5mm}

{\bf Theorem 2}: {CCSs coincide with PSs iff there is a Lindblad
operator $L_\ell$ proportional to each $a_i$ or $a^\dagger_i$, or to
$\sum_i a_i^{\;}$ or $\sum_i a^\dagger_i$ for degenerate modes.}

{\bf Proof}: If no frequency degeneracy occurs, let each $L_\ell$ be
of the form $c_{\ell}^{(i)} a_i$, $ d_{\ell}^{(i)} a^\dagger_i$ for
$c_{\ell}^{(i)}, d_{\ell}^{(i)} \in {\mathbb C}$. From
Eq.~(\ref{wcl2}) and the fact that $(\Delta a_i^\dagger)^2=(\Delta
a_i)^2 + 1$, the rate of purity loss is $\dot{\Pi}_{|\psi\rangle}=
\sum_{\ell,i} (|c_{\ell}^{(i)}|^2 + |d_{\ell}^{(i)}|^2 ) (\Delta
a_i)^2$, up to irrelevant constant terms. CCSs are eigenvectors of
$a_i$, and are therefore PSs. If degeneracies occur, a similar
argument shows that the rate of purity loss contains both terms
proportional to $(\Delta a_i)^2$, and cross terms $\langle a_i^\dagger
a_j \rangle - \langle a_i^\dagger \rangle \langle a_j \rangle$.  Both
are minimized iff ${|\psi \rangle}$ is a CCS.  \hfill{\
\rule{0.5em}{0.5em}}

\vspace*{0.5mm}

Eq.~(\ref{wcl2}) may be used to identify GCSs for bosonic Lie algebras
other than $\mathfrak{h}_3$.  Focus, without loss of generality, on a
single mode. If $\{L_\ell\}$ includes {\em quadratic} bosonic
operators, the corresponding GCSs are squeezed states~\cite{Gil},
resulting from displacements ${\tt D}(\xi,\eta) = \exp(\xi^\ast a^2
-\xi a^{\dagger 2}) {\tt D}(\eta)$, $\xi,\eta \in {\mathbb C}$.  In
general, only a {\em subset} of GCSs emerge as PSs in the WCL, those
that minimize the corresponding scalar uncertainty $(\Delta {\cal
I})^2$: e.g., CCSs are PSs if $a^2$ is among the Lindblad operators,
whereas only $|0\rangle$ is stable if $a^{\dagger\,2}$ and/or
$a^\dagger a$ are also included.

Different scenarios may arise beyond the WCL. Provided that the
effects of the dissipator ${\cal D}$ may still be treated to
first-order ($\tau_R >\tau_S)$, relaxing condition (\ref{wcl}) allows
the $L_\ell(t)$ in Eq.~(\ref{Lt}) to acquire a non-trivial
time-dependence -- causing $\dot{\Pi}_{|\psi\rangle}\not =
\overline{\Pi}_{|\psi\rangle}$ in general. Let us consider the case
where $L_\ell = c_\ell a + d_\ell a^\dagger$, which includes the
quantum Brownian setting of~\cite{Zurek1993}.  As $L_\ell(t) \approx
c_\ell e^{-i\omega t} a +d_\ell e^{i \omega t}a^\dagger$,
Eq.~(\ref{loss}) yields both time-independent contributions and terms
oscillating with frequency $\pm 2\omega t$.  The latter integrate out
if robust PSs are sought by extremizing the average purity loss,
Eq.~(\ref{sieve}) with $\tau \simeq 2\pi/\omega$, leading to
$\overline{\Pi}_{|\psi\rangle} = \sum_\ell (|c_\ell|^2 +|d_\ell|^2)
(\Delta a)^2$, up to constants. Thus, CCSs still emerge as
PSs~\cite{notePSs}.  For intermediate times, the states minimizing
$\dot{\Pi}_{|\psi\rangle}(t)$ may be shown to be squeezed states with
{\em time-dependent} squeezing parameter, in agreement with earlier
results \cite{Old}.

Consider next Markovian evolutions characterized by an irreducible
{\em semisimple} Lie algebra $\mathfrak{g}$.  Paradigmatic examples
are $d$-level systems obeying the quantum optical master
equation~\cite{Scully}. The following result follows from a direct
application of Eq.~(\ref{wcl2}) and Theorem 1:

\vspace*{0.5mm}

{\bf Theorem 3}: GCSs of $\mathfrak{g}$ coincide with PSs {\em if}
$\overline{\Pi}_{|\psi\rangle} = |\lambda|^2 (\Delta {\cal I})^2$.

\vspace*{0.5mm}

Theorem 3 applies, in particular, if the Lindblad operators constitute
an orthonormal basis of $\mathfrak{g}$, up to a global constant
$\lambda \in {\mathbb C}$.  Furthermore, in this case the Lindblad
generator is {\em ergodic} and {\em unital}, implying that all initial
states thermalize~\cite{BP02}, and that the purity decreases
monotonically~\cite{LSA06} -- which puts the first-order approximation
on a firmer ground.  Since a semisimple algebra $\mathfrak{g}$ may be
uniquely expressed as a direct sum of simple algebras,
$\mathfrak{g}=\oplus_u \mathfrak{g}_u$, the requirement of strict
proportionality between $\overline{\Pi}_{|\psi\rangle}$ and $(\Delta
{\cal I})^2$ in Theorem 3 may be weakened by allowing different
proportionality constants for each $\mathfrak{g}_u$.  While
mathematically this leads to a condition which is also {\em necessary}
for all GCSs to be PSs, the basic physical requirement is unchanged:
as minimum uncertainty of GCSs reflects their high degree of symmetry,
enhanced stability against decoherence is only ensured provided that
${\cal D}$ shares, itself, this symmetry. An illustrative situation is
a damped two-level atom with non-radiative dephasing~\cite{Scully}
\begin{eqnarray}
{\cal D}[\rho] &= &  \hspace*{-1mm}  \gamma_1
(\bar{n}+1) ([\sigma_- \rho, \sigma_+] + 
\mbox{h.c.}) +  \label{qome} \\
& + &  \hspace*{-1mm} \gamma_1 \bar{n} ([\sigma_+ \rho, \sigma_-] + 
\mbox{h.c.}) \big ] 
+ \gamma_2 \Big([\sigma_z \rho, \sigma_z] + 
\mbox{h.c.}\Big)\,, \nonumber
\end{eqnarray}
where $\gamma_j>0$, $\bar{n}$ is the thermal photon number, and
$\sigma_{\pm,z}$ are Pauli operators.  PSs are always GCSs in this
case, however arbitrary GCSs are PSs only for high temperature
($\bar{n}+1 \approx \bar{n}$) and $\gamma_2 \approx \bar{n} \gamma_1$.
For higher-dimensional generalizations of (\ref{qome}) PSs need not be
$\mathfrak{su}(2)$-GCSs if the conditions of Theorem 3 are not
fulfilled.

{\em PSs for reducibly represented algebras.}-- States
suffering no purity loss to {\em all} orders in time under Markovian
dynamics have been identified within DFS theory \cite{dfs}.  On one
hand, DFSs are perfect {\em pointer subspaces}.  On the other hand,
DFSs require a {\em reducible} action of $\mathfrak{g}$. Can we relate
DFSs to uncertainty and GCSs?

Consider a reducible representation $\Gamma$ of a compact Lie group
${\cal G}$, with $\Gamma =\oplus_\mu n_\mu \Gamma_\mu$, ${\cal
H}=\oplus_\mu {\cal H}_\mu$ being the associated irrep and state space
decompositions, respectively, and $n_\mu \geq 1$ counting the $\mu$-th
irrep multiplicity.  Theorem 1 allows an explicit characterization of
{\em minimum invariant uncertainty states} to be given starting from
the individual irreps: If $\Lambda_\mu$ is the highest weight of the
$\mu$th irrep, the minimum of $(\Delta {\cal I})^2$ for $\Gamma$ is
achieved by the {\em minimum} highest weight.  Let $|\Lambda\rangle$
be {any} normalized reference state in the $n_{\bar{\mu}}$-dimensional
subspace generated by minimum highest weight vectors.  Then the
minimum-uncertainty manifold may still be formally obtained
through a displacement of $|\Lambda\rangle$ -- as such,
minimum-uncertainty states retain {\em maximum symmetry}, although
they are no longer the unique states with this property.

In the Markovian limit, a DFS subspace ${\cal H}_{\tt DFS}$ is defined
by the property ${\cal D}(|\phi(t)\rangle \langle \phi(t)|)=0$, for
all $|\phi (t)\rangle \in {\cal H}_{\tt DFS}$ and all $t$. If the
corresponding $L_\ell$ close a semisimple $\mathfrak{g}$ and the WCL
is obeyed, ${\cal H}_{\tt DFS}$ consists of the set of states
invariant under ${\cal G}$, that is, the {\em singlet} sector of
$\mathfrak{g}$, $L_\ell |\phi\rangle =0$ for all $\ell$ and
$|\phi\rangle \in {\cal H}_{\tt DFS}$~\cite{dfs}.  Since the singlet
corresponds to the irrep with the minimum highest weight, $(\Delta
{\cal I})^2_{|\phi\rangle}=0$ iff $|\phi\rangle \in {\cal H}_{\tt
DFS}$, thus we have

\vspace*{0.5mm}

{\bf Theorem 4}: Let a DFS-supporting Markovian dynamics be described
by a semisimple Lie algebra $\mathfrak{g}$. Then DFSs are minimum
uncertainty pointer subspaces of $\mathfrak{g}$.

\vspace*{0.5mm}

If GCSs are {\em defined} as minimum uncertainty states, the above
Theorem immediately identifies DFSs with GCSs.  However, GCS
constructions for reducible representations do not retain all
properties which GCSs enjoy for irreps (e.g., minimum-uncertainty GCSs
need not be ground states of Hamiltonians in $\mathfrak{g}$).
Regardless, neither the minimum-uncertainty or the GCS
characterization extend to NSs, which provide the most general pathway
to protecting quantum information~\cite{KLV,Manny06}.  The key
difference brought by the NS notion is the possibility that states in
a {\em factor} ${\cal H}_{\tt NS}$ of a subspace of ${\cal H}$ are
unaffected by ${\cal G}$ -- while allowing {\em arbitrary evolution}
in the full ${\cal H}$.  Thus, NSs are not captured by the definition
of PSs as pure preserved states of $S$ adopted throughout.  Since
genuine NSs live in the multiplicity space of irreps with dimension
higher than one, they necessarily carry {\em non-zero uncertainty}.
We illustrate the DFS-NS comparison in a system of four spin-$1/2$s
with $\mathfrak{g}=\mathfrak{su}(2)$ acting in ${\cal H}=({\mathbb
C}^{2})^{\otimes 4}$ via the total spin representation (collective
decoherence). The irrep decomposition reads $\Gamma^{\otimes 4}_{1/2}
= \Gamma_2 \oplus 3 \Gamma_1 \oplus 2 \Gamma_0 \simeq \Gamma_2 \oplus
(\openone_3 \otimes \Gamma_1) \oplus (\openone_2 \otimes \Gamma_0)$,
$\openone_n$ being the $n$-dimensional identity operator.  The
two-dimensional spin-$0$ sector ${\cal H}_0\equiv {\cal H}_{\tt
DFS}$. A three-dimensional NS lives in the spin-$1$ subspace, ${\cal
H}_1 \simeq {\cal H}_{\tt NS} \otimes {\cal H}_{\tt N}$, where the
noisy factor ${\cal H}_{\tt N}$ is also three-dimensional.  For any
$|\psi_1\rangle \in {\cal H}_1$, $(\Delta {\cal I})^2_{|\psi_1\rangle}
\geq 1$, making the NS invisible to both the invariant uncertainty and
the purity loss functionals.


{\em Conclusions.}-- We have established Lie-algebraic conditions for
the emergence of GCSs as PSs of Markovian quantum evolutions. Although
our analysis rests on invoking purity as a measure of classicality, we
expect that different criteria will agree as long as PSs are well
defined \cite{Dalvit05}.  Notably, spin GCSs have been recently
identified as maximum {\em longevity} states against quantum reference
frame degradation \cite{Bart06}, suggesting the validity of similar
conclusions beyond the Markovian regime.  Yet, as NSs vividly
exemplify, known measures such as purity loss seem too strong a
criterion for sieving robust dynamical features in the presence of the
environment.  What weaker notion of a {\em pointer structure} can
capture the robustness of observable properties of the system, for
instance persistent {\em correlations} in otherwise non-robust states?
From a condensed-matter standpoint, distinguishability of quantum
states according to QFI-related indicators (including topological
quantum numbers and Berry phases) is closely related to
differential-geometric approaches to quantum phase transitions in
matter~\cite{Aligia}.  How does this tie into GCSs and open-system
theory?  Ultimately, answering these questions will be relevant to
synthesize and control novel, stable phases in interacting quantum
systems.

Thanks to H. Barnum, R. Blume-Kohout, C.M. Caves, E.  Knill, A.
Monras, R. Somma, Y. S. Weinstein, P. Zanardi, and W.H. Zurek for
feedback.  L.V.  acknowledges partial support from Constance and
Walter Burke's Special Projects Fund in QIS.  S.B. is supported in
part from {\em la Caixa} fellowship program and the ONR under Grant
No. N00014-03-1-0426.



\begin{thebibliography}{28}


\expandafter\ifx\csname natexlab\endcsname\relax\def\natexlab#1{#1}\fi
\expandafter\ifx\csname bibnamefont\endcsname\relax
  \def\bibnamefont#1{#1}\fi
\expandafter\ifx\csname bibfnamefont\endcsname\relax
  \def\bibfnamefont#1{#1}\fi
\expandafter\ifx\csname citenamefont\endcsname\relax
  \def\citenamefont#1{#1}\fi
\expandafter\ifx\csname url\endcsname\relax
  \def\url#1{\texttt{#1}}\fi
\expandafter\ifx\csname urlprefix\endcsname\relax\def\urlprefix{URL }\fi
\providecommand{\bibinfo}[2]{#2}
\providecommand{\eprint}[2][]{\url{#2}}

\bibitem[{\citenamefont{Schr\"odinger}(1926)}]{Schroedinger26}
\bibinfo{author}{\bibfnamefont{E.}~\bibnamefont{Schr\"odinger}},
  \bibinfo{journal}{Naturwissenshaften} \textbf{\bibinfo{volume}{14}},
  \bibinfo{pages}{664} (\bibinfo{year}{1926}).

\bibitem[{\citenamefont{Glauber}(1963)}]{Roy63}
\bibinfo{author}{\bibfnamefont{R.~J.} \bibnamefont{Glauber}},
  \bibinfo{journal}{Phys. Rev.} \textbf{\bibinfo{volume}{131}},
  \bibinfo{pages}{2766} (\bibinfo{year}{1963}).

\bibitem[{\citenamefont{Sudarshan}(1963)}]{Sud63}
\bibinfo{author}{\bibfnamefont{E.~G.~C.} \bibnamefont{Sudarshan}},
  \bibinfo{journal}{Phys. Rev. Lett.} \textbf{\bibinfo{volume}{10}},
  \bibinfo{pages}{277} (\bibinfo{year}{1963}).

\bibitem[{\citenamefont{Scully and Zubairy}(1997)}]{Scully}
\bibinfo{author}{\bibfnamefont{M.~O.} \bibnamefont{Scully}} \bibnamefont{and}
  \bibinfo{author}{\bibfnamefont{M.~S.} \bibnamefont{Zubairy}},
  \emph{\bibinfo{title}{Quantum Optics}} (\bibinfo{publisher}{Cambridge
  University Press, Cambridge}, \bibinfo{year}{1997}).

\bibitem[{\citenamefont{Zhang et~al.}(1990)\citenamefont{Zhang, Feng, and
  Gilmore}}]{Gil}
\bibinfo{author}{\bibfnamefont{W.~M.} \bibnamefont{Zhang}},
  \bibinfo{author}{\bibfnamefont{D.~H.} \bibnamefont{Feng}}, \bibnamefont{and}
  \bibinfo{author}{\bibfnamefont{R.}~\bibnamefont{Gilmore}},
  \bibinfo{journal}{Rev. Mod. Phys.} \textbf{\bibinfo{volume}{63}},
  \bibinfo{pages}{867} (\bibinfo{year}{1990}).

\bibitem[{\citenamefont{Delbourgo and Fox}(1977)}]{Del}
\bibinfo{author}{\bibfnamefont{R.}~\bibnamefont{Delbourgo}} \bibnamefont{and}
  \bibinfo{author}{\bibfnamefont{J.~F.} \bibnamefont{Fox}},
  \bibinfo{journal}{J. Phys. A} \textbf{\bibinfo{volume}{10}},
  \bibinfo{pages}{L233} (\bibinfo{year}{1977}).

\bibitem[{Tst()}]{Tstab}
\bibinfo{note}{J. R. Klauder, {\tt quant-ph/0110108}; G. D'Ariano, M. Rasetti,
  and M. Vadacchino, J. Phys. A {\bf 18}, 1295 (1985).}

\bibitem[{GE()}]{GE}
\bibinfo{note}{H. Barnum {\em et al.}, Phys. Rev. Lett. {\bf 92}, 107902
  (2004); R. Somma {\em et al.}, Phys. Rev. A {\bf 70}, 042311 (2004).}

\bibitem[{Som()}]{Somma06}
\bibinfo{note}{R. Somma {\em et al.}, Phys. Rev. Lett. {\bf 97}, 190501
  (2006).}

\bibitem[{\citenamefont{Zurek}(2003)}]{Zurek03}
\bibinfo{author}{\bibfnamefont{W.~H.} \bibnamefont{Zurek}},
  \bibinfo{journal}{Rev. Mod. Phys.} \textbf{\bibinfo{volume}{75}},
  \bibinfo{pages}{715} (\bibinfo{year}{2003}).

\bibitem[{Zur()}]{Zurek1993}
\bibinfo{note}{W. H. Zurek, Progr. Theor. Phys. {\bf 89}, 281 (1993); W. H.
  Zurek, S. Habib, and J.-P. Paz, Phys. Rev. Lett. {\bf 70}, 1187 (1993).}

\bibitem[{Old()}]{Old}
\bibinfo{note}{M. Tegmark and H. S. Shapiro, Phys. Rev. E {\bf 50}, 2538
  (1994); M. R. Gallis, Phys. Rev. A {\bf 53}, 655 (1995); Gh.-S. Paraoanu and
  H. Scutaru, Phys. Lett. A {\bf 238}, 219 (1998); A. Isar, Fortschr. Phys.
  {\bf 47}, 855 (1999).}

\bibitem[{\citenamefont{Breuer and Petruccione}(2002)}]{BP02}
\bibinfo{author}{\bibfnamefont{H.-P.} \bibnamefont{Breuer}} \bibnamefont{and}
  \bibinfo{author}{\bibfnamefont{F.}~\bibnamefont{Petruccione}},
  \emph{\bibinfo{title}{The Theory of Open Quantum Systems}}
  (\bibinfo{publisher}{Oxford UP, Oxford}, \bibinfo{year}{2002}).

\bibitem[{dfs()}]{dfs}
\bibinfo{note}{P. Zanardi and M. Rasetti, Phys. Rev. Lett. {\bf 79}, 3306
  (1997); D. A. Lidar, I. L. Chuang, and K. B. Whaley, Phys. Rev. Lett. {\bf
  81}, 2594 (1998).}

\bibitem[{\citenamefont{Knill et~al.}(2000)\citenamefont{Knill, Laflamme, and
  Viola}}]{KLV}
\bibinfo{author}{\bibfnamefont{E.}~\bibnamefont{Knill}},
  \bibinfo{author}{\bibfnamefont{R.}~\bibnamefont{Laflamme}}, \bibnamefont{and}
  \bibinfo{author}{\bibfnamefont{L.}~\bibnamefont{Viola}},
  \bibinfo{journal}{Phys. Rev. Lett.} \textbf{\bibinfo{volume}{84}},
  \bibinfo{pages}{2525} (\bibinfo{year}{2000}).

\bibitem[{cav()}]{caveat}
\bibinfo{note}{${\mathfrak g}$ is a {\em real} Lie algebra of skew-Hermitian
  operators. We identify ${\mathfrak g}$ with its {\em complexification} as
  standard in physics.}

\bibitem[{sym()}]{symmetry}
\bibinfo{note}{That is, maximal ``isotropy subalgebra" $\mathfrak{g}_0=\{g_0
  \in \mathfrak{g}\,| $ $\, g_0 |\Psi_0\rangle= \lambda |\Psi_0\rangle\}$.}

\bibitem[{Are()}]{Arecchi}
\bibinfo{note}{F. T. Arecchi {\em et al.}, Phys. Rev. A {\bf 6}, 2211 (1972).}

\bibitem[{\citenamefont{Braunstein et~al.}(1996)\citenamefont{Braunstein,
  Caves, and Milburn}}]{Braunstein1996}
\bibinfo{author}{\bibfnamefont{S.~L.} \bibnamefont{Braunstein}},
  \bibinfo{author}{\bibfnamefont{C.~M.} \bibnamefont{Caves}}, \bibnamefont{and}
  \bibinfo{author}{\bibfnamefont{G.~J.} \bibnamefont{Milburn}},
  \bibinfo{journal}{Ann. Phys. (N.Y.)} \textbf{\bibinfo{volume}{247}},
  \bibinfo{pages}{135} (\bibinfo{year}{1996}).

\bibitem[{\citenamefont{Hayashi}(2006)}]{hayashi}
\bibinfo{author}{\bibfnamefont{M.}~\bibnamefont{Hayashi}},
  \emph{\bibinfo{title}{Quantum Information: An Introduction}}
  (\bibinfo{publisher}{Springer, Berlin}, \bibinfo{year}{2006}).

\bibitem[{Pro()}]{Proof1}
\bibinfo{note}{$(\Delta {\cal I})^2$ is minimized by GCSs~\cite{Del}. Let
  $\langle \alpha , \beta \rangle$ be the scalar product between two roots.
  Then $(\Delta {\cal I})^2_{|\Lambda, \Lambda\rangle}= \sum_{\alpha \in
  {\mathbb K}} \langle \Lambda, \alpha \rangle \equiv \langle \Lambda, 2\delta
  \rangle$, where ${\mathbb K}$ is the set of positive roots and $2\delta =
  \sum_{\alpha \in {\mathbb K}}\alpha$. Thus, for every $|\psi\rangle$,
  $(\Delta {\cal I})^2_{|\psi\rangle} \geq \langle \Lambda, 2\delta \rangle$.
  The integers $\langle 2 \delta , \alpha_j\rangle/ \langle
  \alpha_j,\alpha_j\rangle\equiv \delta_j$ are the Dynkin coefficients of
  $\delta$. By exploiting the geometry of the Weyl group, $\delta_j=1$ for all
  $j$. Expanding $\Lambda =\sum_j k_j \alpha_j$, and substituting into $\langle
  \Lambda, 2\delta \rangle$ yield the desired lower bound.}

\bibitem[{Lon()}]{Long}
\bibinfo{note}{S. Boixo, L. Viola, and G. Ortiz (in preparation).}

\bibitem[{not()}]{notePSs}
\bibinfo{note}{If $H_S \approx 0$, eigenvectors of $L_\ell$ minimize
  $\dot{\Pi}_{|\psi\rangle}(t)$ -- i.e. {\em position} eigenstates are
  instantaneous PSs~\cite{Zurek1993}. This case, however, is beyond the weak
  dissipation limit we assume.}

\bibitem[{\citenamefont{Lidar et~al.}(2006)\citenamefont{Lidar, Shabani, and
  Alicki}}]{LSA06}
\bibinfo{author}{\bibfnamefont{D.~A.} \bibnamefont{Lidar}},
  \bibinfo{author}{\bibfnamefont{A.}~\bibnamefont{Shabani}}, \bibnamefont{and}
  \bibinfo{author}{\bibfnamefont{R.}~\bibnamefont{Alicki}},
  \bibinfo{journal}{Chem. Phys.} \textbf{\bibinfo{volume}{322}},
  \bibinfo{pages}{82} (\bibinfo{year}{2006}).

\bibitem[{\citenamefont{Knill}(2006)}]{Manny06}
\bibinfo{author}{\bibfnamefont{E.}~\bibnamefont{Knill}},
  \bibinfo{journal}{Phys. Rev. A} \textbf{\bibinfo{volume}{74}},
  \bibinfo{pages}{042301} (\bibinfo{year}{2006}).

\bibitem[{\citenamefont{Dalvit et~al.}(2005)\citenamefont{Dalvit, Dziarmaga,
  and Zurek}}]{Dalvit05}
\bibinfo{author}{\bibfnamefont{D.~A.~R.} \bibnamefont{Dalvit}},
  \bibinfo{author}{\bibfnamefont{J.}~\bibnamefont{Dziarmaga}},
  \bibnamefont{and} \bibinfo{author}{\bibfnamefont{W.~H.} \bibnamefont{Zurek}},
  \bibinfo{journal}{Phys. Rev. A} \textbf{\bibinfo{volume}{72}},
  \bibinfo{pages}{062101} (\bibinfo{year}{2005}).

\bibitem[{Bar()}]{Bart06}
\bibinfo{note}{S. D. Bartlett {\em et al.}, New J. Phys. {\bf 8}, 58 (2006).}

\bibitem[{Ali()}]{Aligia} \bibinfo{note}{A. A. Aligia and G. Ortiz,
Phys. Rev. Lett. {\bf 82}, 2560 (1999); G. Ortiz and A. A. Aligia,
Phys. Stat. Sol. (b) {\bf 220}, 737 (2000). Recently, similar
conclusions have been reached e.g. in P. Zanardi, P. Giorda, M. Cozzini,
{\tt quant-ph/0701061}. }

\end{thebibliography}

\vspace*{-3mm}

\end{document}